\documentclass[11pt,prd,floats,eqsecnum,nofootinbib]{revtex4}

\def\Rbol{{\stackrel{\circ}{R}}{}}
\def\Sbol{{\stackrel{\circ}{\mathcal S}}{}}
\def\ombol{{\stackrel{\circ}{\omega}}{}}
\def\tref{h_{(\rm r)}}
\def\Lbol{{\stackrel{\circ}{\mathcal L}}{}}
\def\Gammaw{{\stackrel{\bullet}{\Gamma}}{}}

\def\tw{{\stackrel{\bullet}{t}}{}}
\def\Lw{{\stackrel{~\bullet}{\mathcal L}}{}}
\def\Tw{{\stackrel{\bullet}{T}}{}}

\def\Kw{{\stackrel{\bullet}{K}}{}}
\def\om{{\omega}{}}
\def\omw{{\stackrel{\bullet}{\omega}}{}}
\def\Sw{{\stackrel{\bullet}{S}}{}}
\def\actionw{{\stackrel{\bullet}{\mathcal S}}{}}

\begin{document}
\title{Spin Connection and Renormalization of Teleparallel Action}
\author{Martin Kr\v s\v s\'ak}
\email{krssak@ift.unesp.br}
\author{J. G. Pereira}
\email{jpereira@ift.unesp.br}
\affiliation{{\it Instituto de F\'isica Te\'orica, Universidade Estadual Paulista}\\
{\it R. Dr. Bento Teobaldo Ferraz 271, 01140-070 S\~ao Paulo SP, Brazil}}

\begin{abstract} 
In general relativity, inertia and gravitation are both included in the Levi-Civita connection. As a consequence, the gravitational action, as well as the corresponding energy-momentum density, are in general contaminated by spurious contributions coming from inertial effects. In teleparallel gravity, on the other hand, because the spin connection represents inertial effects only, it is possible to separate inertia from gravitation. Relying on this property, it is shown that to each tetrad there is naturally associated a spin connection that {\em locally removes the inertial effects} from the action. The use of the appropriate spin connection can be viewed as a renormalization process in the sense that the computation of energy and momentum naturally yields the physically relevant values. A self-consistent method for solving field equations and determining the appropriate spin connection is presented.
\end{abstract}
\maketitle

\section{Introduction} 

The search for a local energy-momentum density for gravity is one of the oldest problem of general relativity. Its difficulty is usually understood to be a consequence of the  equivalence principle, which locally identifies inertial with gravitational effects.
To illustrate this point, let us consider two observers in a gravitational field of some massive object: one is in free-fall and the other is kept at a fixed distance from the object. For the first observer gravity seems to be switched off, and she/he naturally assigns a zero energy-momentum density to the gravitational field. The other observer experiences a gravitational pull and assigns a non-zero energy-momentum density to the same gravitational field. This does not mean that it is impossible to define unequivocally an energy-momentum density for gravity. Rather, it means simply that, in order to define it, we must be able to separate inertial from gravitational effects.

In the context of general relativity, both inertial and gravitational effects are encoded in the Levi-Civita connection, and cannot be separated. For this reason, any complex defining the gravitational energy-momentum in this theory density will include the contribution coming from inertial effects, and will consequently be a non-covariant quantity---that is, a pseudotensor \cite{Landau:1982dva}. An attempt to circumvent this problem is the so-called quasi-local approach, in which one defines the energy-momentum associated with a region of spacetime $\mathcal{M}$ with boundary $\partial\mathcal{M}$. The full gravitational action in this case is given by \cite{York:1972sj,Gibbons:1976ue}\footnote{We denote all quantities related to general relativity with a $``\circ"$ over the quantity.}
\begin{equation}
\Sbol(g) = \int_{\mathcal{M}} \Lbol_{EH} + \int_{\partial\mathcal{M}}  \Lbol_{GHY},
\label{actbol}
\end{equation}
where $\Lbol_{EH}$ is the Einstein-Hilbert Lagrangian and $\Lbol_{GHY}$ is the Gibbons-Hawking-York boundary term. The action (\ref{actbol}) was originally used in the framework of Euclidean gravity, where many important results about black hole thermodynamics were obtained \cite{Gibbons:1976ue}. It was then used in the Hamiltonian formalism, where the role of boundary terms was investigated \cite{Regge:1974zd,Hawking:1995fd}, as well as in the quasi-local definition of gravitational energy-momentum tensor \cite{Brown:1992br}. 

The problem of  the  action (\ref{actbol}) is that it suffers from IR divergences: it diverges when the boundary is taken to the infinity. This happens even for a flat metric written in a general coordinate system, in which case the action (\ref{actbol}) represents inertial effects only. It is then clear that such divergences need to be removed to yield physically sensible results. These divergences are closely related to the problem of the asymptotic limit of the Lagrangian density: if the Lagrangian density does not vanish at infinity, we obtain a divergent action. Notice however that, since any physical field must vanish far enough away from the source, the purely gravitational action is expect to be free of IR divergences.

On the other hand, the energy associated to the inertial effects does not vanish at infinity. This can be seen already in classical Newtonian physics: if we consider an observer in the gravitational field of some planet, and we allow such observer to rotate (i.e. include some inertial effects), this observer would measure that the planet has some rotational energy in respect to him. As we increase a distance between the observer and the planet, this energy would increase and be unavoidably divergent. These results constitute a clear evidence that the inertial effects are  responsible for the IR divergences of the action.

In general relativity, this problem is typically addressed through the so-called ``background subtraction" method, proposed by Gibbons and Hawking \cite{Gibbons:1976ue},  which
considers a reference spacetime $\mathcal{M}_{\rm ref}$ related to a flat metric $g_{\rm ref}$, which is isometrically embedded in the general spacetime $\mathcal{M}$. The physically renormalized action is then defined as the difference 
\begin{equation}
\Sbol_{\rm ren}=\Sbol(g)-\Sbol(g_{\rm ref}).
\label{subt}
\end{equation}  
The underlying idea is that the divergences  are  fully encoded in the reference spacetime $\mathcal{M}_{\rm ref}$, and the subtraction (\ref{subt}) has the effect of removing all divergent contributions from the action. The action for the reference spacetime represents just inertial effects, since it is a flat spacetime, where gravity is absent. Subtraction (\ref{subt}) can thus be understood as a removal of the inertial effects from the action.

In the Gibbons-Hawking renormalization process, the divergences are removed only quasi-locally---that is to say, are removed from the action as an integral over the whole spacetime. As a consequence, it is not always possible to construct an open family of the embeddings of the reference spacetime around a given solution. For this reason, the variational principle associated to the renormalized gravitational action (\ref{subt}) is not always well-defined \cite{Mann:2005yr}. On the other hand, owing to the fact that in teleparallel gravity \cite{Aldrovandi:2013wha} the spin connection represents inertial effects only, it becomes possible to locally separate inertial and gravitational effects---something impossible to achieve in general relativity. As a consequence, one is able to locally remove the inertial effects from the action, yielding in this way a purely gravitational action. Of course, the same procedure can be used to locally remove the inertial effects from the gravitational energy-momentum density, which yields a local notion for energy and momentum.

Working in the context of teleparallel gravity, the purpose of this paper is to investigate the problem of defining a purely gravitational action, to the exclusion of inertial effects, which will be free of divergences. The corresponding gravitational energy-momentum current will also be free of the spurious inertial effects, and consequently will always yield the physical result. We are going to proceed as follows. In Section~\ref{sec1} we briefly introduce the fundamentals of teleparallel gravity. In Section~\ref{sec4} we review the local Lorentz invariance of teleparallel gravity, and introduce the teleparallel spin connection. We then show that to each tetrad, there is a naturally associated spin connection. A method for retrieving such spin connection from the tetrad is provided, which can be considered the main result of the paper. In Section~\ref{sec6}, using such method, we develop a self-consistent approach for solving the field equations of teleparallel gravity. In Section~\ref{sec5} we show that associating the appropriate spin connection to a given tetrad removes the inertial effects from the action. Finding the appropriate spin connection, therefore, can be viewed as a renormalization process. We illustrate this property for the cases of Schwarzschild and Kerr solutions. In Section~\ref{PureTetra}, for the sake of comparison, we briefly comment on the so-called pure tetrad teleparallel gravity, and finally in Section~\ref{conclu} we summarize and discuss the results obtained.

\section{Fundamentals of Teleparallel Gravity
\label{sec1}}

From now on, we will use the Greek alphabet ($\mu, \nu, \rho. \dots$) to denote spacetime indices and the Latin alphabet ($a, b, c, \dots $) to denote tangent-space indices. At every point of spacetime, the tetrad defines a vector basis for the corresponding Minkowski tangent space. From a physical viewpoint, the  tetrad $h^a_{\ \mu}$ represents a local frame of reference and is related to the spacetime metric by
\begin{equation}
g_{\mu\nu}=\eta_{ab} h^a_{\ \mu} h^b_{\ \nu},
\label{metdef}
\end{equation}
where $\eta_{ab}=\text{diag} (1,-1,-1,-1)$ is the Minkowski metric of the tangent space.

From a physical viewpoint, there are two spin connections of special interest. The Levi-Civita spin connection $\ombol^a_{\ b \mu}$, which is the relevant connection of general relativity, is the unique connection with vanishing torsion. Analogously, there is a unique connection with vanishing curvature: it is the so-called \textit{Weitzenb\"ock} connection $\omw^a_{\ b \mu}$, the relevant connection of the \textit{teleparallel equivalent of general relativity}, or teleparallel gravity for short\footnote{We denote all quantities related to teleparallel gravity with a $``\bullet"$ over the quantity.}. These two connections are related by the Ricci theorem  \cite{KoNu}
\begin{equation}
\omw^a_{\ b \mu}=\ombol^a_{\ b \mu}+\Kw^a_{\ b \mu},
\label{decomp}
\end{equation}
where
\begin{equation}
\Kw^{a}_{\  b\mu}=\frac{1}{2}
\left(
\Tw^{\ a}_{\mu \ b}
+\Tw^{\ a}_{b \ \mu}
-\Tw^{a}_{\  b\mu}
\right),
\label{contortion}
\end{equation}
is the contortion tensor and $T^\rho_{\ \mu\nu}$ is the torsion tensor. 

The Lagrangian of teleparallel gravity is then written in the form \cite{Aldrovandi:2013wha}
\begin{equation}
\Lw=\frac{h}{4 \kappa} \Tw^\alpha_{\ \rho\sigma}\Sw_\alpha^{\ \rho\sigma}, \label{lagtot}
\end{equation}
where $h=\det h^a_{\ \mu}$, $\kappa=8\pi G$ is the gravitational constant, and 
\begin{equation}
\Sw^{\rho\mu\nu}=-\Sw^{\rho\nu\mu}=\Kw^{\mu\nu\rho}
-g^{\rho\nu}\Tw^{ \mu}
+g^{\rho\mu}\Tw^{ \nu},
\label{superp}
\end{equation}
is the superpotential, with $\Tw^{\mu}=\Tw^{\nu \mu}{}_{\nu}$ the vector torsion. Using Ricci's theorem (\ref{decomp}), it is straightforward to show that the teleparallel Lagrangian (\ref{lagtot}) differs from the Einstein-Hilbert Lagrangian of general relativity by a total divergence:
\begin{equation}
\Lw =\Lbol_{EH}-\partial_\mu \left(\frac{h}{\kappa} \, \Tw^\mu \right).
\end{equation}
The teleparallel equations of motion are obtained by varying the Lagrangian (\ref{lagtot}) with respect to the tetrad $h_a^{\ \mu}$. In vacuum, we find that the spacetime-indexed equations are given by
\begin{equation}
\partial_\sigma \Big(h \Sw_\mu^{\ \rho\sigma} \Big)+
\kappa  h \tw_\mu^{\ \rho}=0, \label{steq}
\end{equation}
where
\begin{equation}
h \tw_\mu^{\ \rho}=   \frac{1}{\kappa} h  \Gammaw^\alpha_{\ \sigma\mu}\Sw_{\alpha}^{\ \sigma\rho}+\delta_\mu^{\ \rho} \Lw, \label{ptensor}
\end{equation}
is the energy-momentum pseudotensor, and $\Gammaw^\alpha_{\ \sigma\mu}$ is the linear connection 
\begin{equation}
\Gammaw^\rho_{\ \nu\mu}=h_a^{\ \rho}\partial_\mu h^a_{\ \nu}+
h_a^{\ \rho} \omw^a_{\ b\mu}h^b_{\ \nu}.
\end{equation}
Of course, on account of the equivalence between the Lagrangians, the teleparallel field equations are found to be equivalent to the Einstein's equations:
\begin{equation}
\partial_\sigma \Big(h \Sw_
\mu^{\ \rho\sigma} \Big)+
 \kappa h \tw_\mu^{\ \rho}\equiv h \left(\Rbol_\mu^{\ \rho}-\frac{1}{2}\delta_\mu^\rho\Rbol\right).\label{feeq}
\end{equation}
Since the Ricci tensor is symmetric, the expression on the left side of (\ref{feeq}) must be symmetric too. As a consequence, we have only ten independent field equations.

Due to the anti-symmetry of the superpotential (\ref{superp}) in the last two indices, it follows that the energy-momentum pseudotensor is conserved in the ordinary sense
\begin{equation}
\partial_\rho\Big(h \tw_\mu^{\ \rho} \Big)=0.
\end{equation} 
As is well-known, such conservation law yields a conserved charge---that is, a time-conserved quantity. Let us consider a spacelike slice $\Sigma$ with $u^\mu$ a timelike unit vector normal to $\Sigma$. Using polar coordinates $(r,\theta,\phi)$ on $\Sigma$, and denoting the unit vector that is normal to the surface of constant radial distance by $n^\mu$, the conserved charge is found to be the four-momentum 
\begin{equation}
P_\mu=\int_\Sigma  d^3x \left(h u_\nu \tw_\mu^{\ \nu} \right) =-\frac{1}{\kappa}
\int_{\partial \Sigma}d\phi d\theta \Big(h u_\nu n_\rho \Sw_\mu^{\  \nu\rho} \Big), \label{dens}
\end{equation}
where we have used Stoke's theorem and the equations of motion (\ref{steq}) in the second equality.

\section{Local Lorentz Transformations and Spin Connections}
\label{sec4}

A local Lorentz transformation is a transformation of the tangent-space coordinates
\begin{equation}
x'^{a}=\Lambda^a{}_{b} x^{b},
\end{equation}
where $\Lambda^a_{\ b}=\Lambda^a_{\ b}(x)$ are point-dependent elements of the Lorentz group. Under such transformation the tetrad changes according to
\begin{equation}
h'^{a}_{\ \mu}=\Lambda^a{}_{b}h^{b}_{\ \mu},
\label{tettransf}
\end{equation}
whereas the spin connection undergoes the transformation
\begin{equation}
\om'{}^a_{\ b\mu}=\Lambda^a{}_{c} \om{}^c_{\ d\mu}\Lambda_b^{\ d}+\Lambda^a_{\ c} \partial_\mu \Lambda_b{}^{c},
\label{spintransf}
\end{equation}
with $\Lambda_b{}^{d}$ the inverse matrix to $\Lambda^b{}_{d}$. A local Lorentz transformation, therefore, amounts to simultaneously transform the tetrad (\ref{tettransf}) and the spin connection (\ref{spintransf}). 

The spin connection of general relativity represents both gravitation and inertial effects. On the other hand, the spin connection of teleparallel gravity represents inertial effects only. This means that there exists a class of frames---called proper frames---in which it vanishes: $\omw^{a}{}_{b\mu}=0$. In a general class of frames, therefore, according to the transformation (\ref{spintransf}), it will assume the form \cite{Aldrovandi:2013wha}
\begin{equation}
\omw^{a}{}_{b\mu}=\Lambda^{a}{}_{c} \, \partial_{\mu} \Lambda_{b}{}^{c}.
\label{InerConn}
\end{equation}
This expression can also be obtained by considering teleparallel gravity as ``embedded" in more general gauge theories, like for example the metric-affine theory \cite{Obukhov:2002tm} or the Poincar\'e gauge theory \cite{HB}. Starting with such general theories, and introducing via Lagrange multipliers the condition of vanishing curvature and non-metricity in the former case, and the condition of vanishing curvature in the later case, one ends up again with the inertial connection (\ref{InerConn}).

As a gauge theory for the translational group \cite{Hayashi:1967se}, the tetrad in teleparallel gravity has always the form \cite{Aldrovandi:2013wha}
\begin{equation}
h^a_{\ \mu}= \partial_\mu x^a + \omw^a_{\ b\mu} x^b + B^a_{\ \mu}. \label{gaugetet}
\end{equation}
Now, the teleparallel field equations are concerned with gravity only: they determine the gravitational potential $B^a_{\ \mu}$ only. In other words, the teleparallel spin connection is not determined by the field equations. This means that the teleparallel field equations are able to determine the tetrad up to a local Lorentz transformation \cite{Obukhov:2006sk}.\footnote{Recall that in the tetrad formulation of general relativity, Einstein equation determines the tetrad up to a local Lorentz transformation \cite{Weinberg}. Due to the equivalence (\ref{feeq}) between the field equations, it is natural to expect that the same holds in teleparallel gravity.}

Then comes the problem: any real computation presupposes a given frame, or tetrad. If this tetrad represents a proper frame, the associated spin connection vanishes: $\{\tilde h^a{}_\mu, 0 \}$. In any other class of frames related to the proper frames by a local Lorentz transformation, the spin connection will be non-vanishing, which means that there are infinitely many pairs $\{h^a{}_\mu, \omw^{a}{}_{b\mu} \}$. Of course, for consistency reasons, the same spin connection must be used in all covariant derivatives. Therefore, we need to provide a mechanism to determine the spin connection associated to a given tetrad.

To begin with, we note that the tetrad (\ref{gaugetet}) satisfies the teleparallel field equations for any spin connection $\omw^a_{\ b\mu}$. Therefore, any given tetrad will also be a solution to the field equations.\footnote{In Section~\ref{sec6} we discuss in details how to solve the teleparallel field equations.} Our main task is then to retrieve the spin connection from the tetrad. We start by considering a ``reference tetrad" $\tref^{\;a}{}_{\mu}$, in which gravity is switched-off. In such tetrad, the translational potential $B^a_{\ \mu}$ vanishes and the reference tetrad can be written as
\begin{equation}
\tref^{\;a}{}_{\mu}=\partial_\mu x^a + \omw^a_{\ b\mu}x^b. \label{inerdecomp}
\end{equation} 
Substituting the decomposition (\ref{inerdecomp}) into the definition of the torsion tensor, it is easy to check that the torsion tensor for the reference tetrad  vanishes identically
\begin{equation}
\Tw^a_{\ \mu\nu}(\tref^{\;a}{}_{\mu}, \omw^a_{\ b\mu})= 0. \label{torzero}
\end{equation}
As we have said, the decomposition (\ref{gaugetet}) is always implicit, and hence we do not immediately know how to set $B^a_{\ \mu}$ equal to zero. However, given a general tetrad $h^a_{\ \mu}$, we can obtain the reference tetrad in which gravity is switched-off by setting some parameter that controls a strength of gravity to zero. The obvious choice is the gravitational constant $G$. Hence, in practice, the reference tetrad is obtained as
\begin{equation}
\tref^{\;a}{}_{\mu}\equiv\left. h^a_{\ \mu}\right|_{G\rightarrow 0}. \label{reftet}
\end{equation}

We define the coefficient  of anholonomy $f^c{}_{a b}$ of the tetrad $h^a=h^a_{\ \mu}dx^\mu$ as
\begin{equation}
[h_a, h_b] = f^c{}_{a b} \, h_c.
\end{equation}
As a simple computation shows, its explicit form is
\begin{equation}
f^c{}_{a b} = h_a{}^{\mu} h_b{}^{\nu} (\partial_\nu
h^c{}_{\mu} - \partial_\mu h^c{}_{\nu} ).
\label{fcab}
\end{equation}
In terms of $f^c{}_{a b}$, torsion can be written as
\begin{equation}
\Tw^a{}_{bc} = - f^a{}_{bc} + (\omw^a{}_{cb} - \omw^a{}_{bc}).
\label{AnhoTor}
\end{equation}
Condition (\ref{torzero}) for the reference tetrad assumes then the form
\begin{equation}
\Tw^a{}_{bc} (\tref^{\;a}{}_{ \mu},\omw^a_{\ b\mu}) = \omw^a{}_{cb} - \omw^a{}_{bc} - f^a{}_{bc}(\tref) =  0,
\label{toree}
\end{equation}
with $f^a{}_{bc}(h_{(\rm r)})$ the coefficient of anholonomy (\ref{fcab}) of the reference tetrad $h_{(\rm r)}^{\;a}{}_{\mu}$. Using (\ref{toree}) for three different combination of indices, we can solve for the spin connection:
\begin{equation}
\omw^a{}_{b\mu} = \frac{1}{2} h_{(\rm r)}^{\;c}{}_{\mu} \Big[f_b{}^a{}_c(h_{(\rm r)}) + f_c{}^a{}_b(h_{(\rm r)}) - f^a{}_{bc}(h_{(\rm r)}) \Big]. 
\label{regconsolution}
\end{equation}
This is the  teleparallel spin connection naturally associated to the reference tetrad $h_{(\rm r)}^{\;a}{}_{\mu}$. Since the reference tetrad $h_{(\rm r)}^{\;a}{}_{\mu}$ and the original tetrad $h^a_{\ \mu}$ differs only in their gravitational content, while the inertial effects are equally present in both of them, the spin connection (\ref{regconsolution}) is the teleparallel spin connection naturally associated with the tetrad $h^a{}_\mu$ as well.

One should note that the inertial spin connection (\ref{regconsolution}) coincides with the Levi-Civita connection for the reference tetrad
\begin{equation}
\omw^a{}_{b\mu} = \ombol^a_{\ b\mu}(\tref). \label{leveq}
\end{equation}
Owing to this relation, our method is found to lead to the same results as  
the one of Refs.~\cite{Obukhov:2006sk,Lucas:2009nq}, where the authors used the asymptotic limit of the Levi-Civita connection. In the case of asymptotically flat spacetimes, it is obvious that such an approach defines a flat connection, but it is rather \textit{ad-hoc}. Here, we have developed a completely new approach to this problem, which gives rise to a whole new method of eliminating the spurious contribution from inertial effects. It should be noted that, in spite of the relation (\ref{leveq}), our method is conceptually independent of the Levi-Civita connection---and consequently of general relativity.

\section{A Note on Solution of the Field Equations\label{sec6}}

Let us briefly discuss the problem of solving the teleparallel field equations.  A simple analysis shows that the problem is defined circularly: the spin connection is determined using (\ref{regconsolution}), which requires the reference tetrad obtained from the solution of the field equations (\ref{reftet}). However, the field equations require the knowledge of torsion, which is a function of both tetrad and spin connection. We now explain how to avoid this difficulty and  provide a self-consistent method to solve field equations in the framework of teleparallel gravity. 

It is important to understand that the starting point of any calculation is the ansatz tetrad, which is chosen in such a way that it reproduces the ansatz metric, which is in turn given by the symmetry of the problem under consideration. This choice is non-unique in the sense that there are infinitely many tetrads leading to the same metric, all of them related by local Lorentz transformations. In practice, we  choose the simplest tetrad. For example, in the case of spherical static spacetime, the most natural choice is the diagonal tetrad in the spherical coordinate system. So far, the situation is similar to the tetrad formulation of general relativity. 

In general relativity, the Levi-Civita spin connection is fully determined by the tetrad, which allows us to solve Einstein equations using an ansatz tetrad only. However,  in teleparallel gravity, the spin connection is needed as well to find torsion and solve the field equations. 
The key observation here is that the field equations do not depend on the spin connection, as  discussed in the previous section. This allows us to solve the field equations using an arbitrary spin connection of the form (\ref{InerConn}). The most natural choice is to consider a vanishing spin connection, which is trivially of the right form (\ref{InerConn})---but in principle any other spin connection of this form could be chosen. Using this property, we determine the tetrad from the field equations, and we can then proceed to find the associated spin connection  using (\ref{reftet}) and (\ref{regconsolution}).

The method described here allows us to solve the field equations and determine the spin connection in a self-consistent way. Alternatively, it is possible to use the equivalence with general relativity (\ref{feeq}), and calculate the spin connection associated to the tetrad---which is already known to solve Einstein equations. In Section~\ref{sec5} we illustrate the use of both methods in the case of the Schwarzchild solution.

\section{Renormalization of the Action and Energy-Momentum \label{sec5}}

We have shown that in teleparallel gravity, to each tetrad there is naturally associated a corresponding spin connection. Then, we have developed a method to retrieve this spin connection from the tetrad. Now, we would like to show that from a physical viewpoint the role of the spin connection is to remove the inertial effects from the action, providing in this way a purely gravitational action which, as discussed in the Introduction, is expected to be finite.

We start by considering an action for the reference tetrad (\ref{reftet}), which represents only inertial effects. If we naively associate a vanishing spin connection to the reference tetrad $\tref^{\;a}{}_{\mu}$, the gravitational action assumes the form 
\begin{equation}
\Sw(\tref^{\;a}{}_\mu, 0 )=\int_\mathcal{M}\Lw(\tref^{\;a}{}_\mu, 0 ).
\label{reftetact}
\end{equation}
In general this action does not vanish, and is even typically divergent. The reason for this result is that it is an action for inertial effects. If instead of a vanishing spin connection we choose the appropriate spin connection (\ref{regconsolution}), due to (\ref{torzero}) we have
\begin{equation}
\Sw(\tref^{\;a}{}_\mu, \omw^a_{\ b\mu} )=0. \label{srefac}
\end{equation}
We see in this way that the role of spin connection $\omw^a_{\ b\mu} $ is to remove all inertial effects of the action, in such a way that it now vanishes---as it should because it represents only inertial effects. One should note that the spin connection removes the inertial effects not from the whole action integral (\ref{srefac}), but locally at each point of the space-time. This is clear from the fact that, not only the action, but also the Lagrangian itself vanishes: $\Lw(\tref^{\;a}{}_\mu, \omw^a_{\ b\mu} ) = 0$. 

From the viewpoint of inertial effects, the full and reference tetrads are equivalent. This means that the spin connection is able to remove the inertial contributions from the full action as well. This yields an action that represents gravitational effects only. Since the inertial effects are responsible for causing the divergences, the purely gravitational action
\begin{equation}
\actionw_{\rm ren}=\int_\mathcal{M} \Lw(h^a{}_{\mu},\omw^a{}_{b\mu}), \label{actren}
\end{equation}
can be viewed as a \textit{renormalized action}. The process of finding the appropriate spin connection to a given tetrad can thus be viewed as a renormalization process. Conceptually, this resembles the Gibbons-Hawking renormalization method that we have discussed in the Introduction. However, one should note that the inertial effects are removed locally at each point of spacetime---and not from the whole integral, as it happens in the Gibbons-Hawking formalism. As a consequence, instead of quasi-local, the energy and momentum densities in teleparallel gravity can be defined locally.

In what follows we illustrate our method of computing the appropriate spin connection and demonstrate its effect on action and energy-momentum density in two cases: the Schwarzschild and the Kerr solutions. 

\subsection{Schwarzschild Solution \label{schw}}

The simplest non-trivial example of the gravitational field is the spherically symmetric Schwarzschild solution, whose metric has the form 
\begin{equation}
ds^2=f(r)dt^2-\frac{1}{f(r)}dr^2-r^2d\Omega^2.
\label{schwmet}
\end{equation}
As is well-known, there are infinitely many tetrads that yield the above metric. As an example, let us consider diagonal tetrad
\begin{equation}
h^a_{\ \mu}=\text{diag}\left( \sqrt{f(r)},
1/\sqrt{f(r)},r, r\sin \theta \right). 
\label{tetdiag}
\end{equation}
To find a solution for the function $f(r)$, we can proceed in two different ways. Firstly, we can refer to the result of general relativity, where
\begin{equation}
f(r)=1-2m/r, \label{alphasol}
\end{equation}
with $m=GM$ is well known. However, it is equally possible to obtain this solution in the framework of teleparallel gravity following the method described in Section~\ref{sec6}. We can set the spin connection to zero, and solve the field equations for the diagonal tetrad (\ref{tetdiag}), which leads to the solution (\ref{alphasol}) \cite{Zet:2003aa}. But, we can check that this zero spin connection is not a correct inertial spin connection associated with the diagonal  tetrad (\ref{tetdiag}), since the teleparallel action is given by
\begin{equation}
\actionw(h^a_{\ \mu},0)=\frac{1}{\kappa}\int_\mathcal{M}d^4 x \sin\theta. \label{schwactnon}
\end{equation}
As a simple inspection shows, it is divergent, which means inertial effects are still included in this action, and hence they were not removed by the spin connection.
We then use the method developed in this paper to find the spin connection associated to the tetrad (\ref{tetdiag}). The starting point is to define the reference tetrad
\begin{equation}
h_{(\rm r)}^{\;a}{}_{\mu} \equiv
\left. h^a_{\ \mu}\right|_{G=0}=
\text{diag}\left( 1, 1, r, r\sin \theta \right).
\end{equation}
Using (\ref{regconsolution}), we find that the non-vanishing components of the spin connection are 
\begin{equation}
\omw^{\hat{1}}_{\ \hat{2}\theta}=-\omw^{\hat{2}}_{\ \hat{1}\theta}=-1, \quad 
\omw^{\hat{1}}_{\ \hat{3}\phi}=-\omw^{\hat{3}}_{\ \hat{1}\phi}=-\sin\theta, \quad
\omw^{\hat{2}}_{\ \hat{3}\phi}=-\omw^{\hat{3}}_{\ \hat{2}\phi}=-\cos\theta. \label{spschw}
\end{equation}
The corresponding renormalized action is found to be
\begin{equation}
\actionw_{\rm ren}(h^a{}_{\mu},\omw^a_{\ b\mu})=\frac{2}{\kappa}
\int_\mathcal{M}d^4 x \left[1+\frac{(m-r)}{r\sqrt{f}} \right]
\sin\theta = \int dt M. \label{schwrenact}
\end{equation}
As expected, it is finite and free of divergences. 
Using (\ref{dens}) we can check that the energy-momentum density is 
\begin{equation}
P_\mu =(M,0,0,0), \label{schwden}
\end{equation}
which is the physically relevant result. We can see that the renormalized action (\ref{schwrenact}) can be then directly interpreted as a time integral of the total energy of the Schwarzchild black hole.

\subsection{Kerr Solution}
In Boyer-Lindquist coordinates $r,\theta,\phi$, the Kerr metric is written as
\begin{equation}
ds^2 = dt^2 - \frac{\Sigma^2}{\Delta} \, dr^2 - (r^2+a^2) \sin^2\theta \, d\phi^2 -
\Sigma^2 \, d\theta^2-\frac{2M r}{\Sigma^2} \, (dt - a\sin^2\theta \, d\phi)^2,
\label{metric}
\end{equation}
where
\[
\Sigma=r^2+a^2\cos^2\theta \quad \mbox{and} \quad \Delta=r^2+a^2-2 m r,
\]
with $a$ the angular momentum per unity mass. A particular tetrad yielding this metric is \cite{Lucas:2009nq}
\begin{equation}
h^a_{\ \mu}=
\left( \begin{array}{cccc}
\frac{\sqrt{\Delta\Sigma}}{A} & 0 & 0 & 0\\ 
0 & \sqrt{\frac{\Sigma}{\Delta}} & 0 \\
0 & 0 & \sqrt{\Sigma} & 0 \\
-\frac{2 a m r}{A\sqrt{\Sigma}}\sin\theta & 0 & 0 & \frac{A}{\sqrt{\Sigma}}\sin\theta
\end{array}
\right),
\label{Kerrtet}
\end{equation}
where 
\begin{equation}
A^2= \Delta\Sigma +2 m r (r^2+a^2).
\end{equation}
Similarly as in the Schwarzchild case, we can write naively the action 
\begin{equation}
\actionw(h^a_{\ \mu},0)=
\frac{1}{\kappa}
\int_\mathcal{M}d^4 x
\frac{r^4+a^2\cos^2\theta(4Mr+a^2\cos^2\theta)}{\Sigma^2}\sin\theta. \label{sactnr}
\end{equation}
To demonstrate that this quantity is divergent, we find that the asymptotic expansion of the Lagrangian is 
\begin{equation}
\Lw(h^a_{\ \mu},0)=
\frac{1}{\kappa}\sin\theta + O\left(
\frac{1}{r^2}
\right).
\end{equation}
Since the leading term in expansion is constant, the action integral (\ref{sactnr}) is consequently IR-divergent.
Our task now is to find the spin connection associated with the above tetrad that  renormalizes the action. To this end, we define first the reference tetrad according to (\ref{reftet}),
and then by using Eq.~(\ref{regconsolution}) we find the non-vanishing components of the spin connection:
\begin{eqnarray}
&& \omw^{\hat{1}}_{\ \hat{2}r} = -\omw^{\hat{2}}_{\ \hat{1}r}=
-\frac {a^2 \cos\theta \sin\theta}
{\sqrt{r^2+a^2}\,\Sigma},\, \quad 
\omw^{\hat{1}}_{\ \hat{2}\theta} = -\omw^{\hat{2}}_{\ \hat{1}\theta}=
-\frac {r\sqrt{r^2+a^2}} {\Sigma}, \nonumber \\
&& {} \\
&&\omw^{\hat{1}}_{\ \hat{3}\phi} = -\omw^{\hat{3}}_{\ \hat{1}\phi}=
-\frac {r \sin\theta}
{\sqrt{\Sigma}}, \qquad \quad ~~ 
\omw^{\hat{2}}_{\ \hat{3}\phi} = -\omw^{\hat{3}}_{\ \hat{2}\phi}=
-\frac {\sqrt{r^2+a^2} \, \cos\theta} {\sqrt{\Sigma}} \nonumber.
\end{eqnarray}
The explicit expression for the renormalized action is too lengthy to be written here, but we can find the leading term in an asymptotic expansion of the Lagrangian:
\begin{equation}
\Lw(h^a_{\ \mu},\omw^a_{\ b\mu})=
-\frac{m^2}{\kappa} \frac{\sin\theta}{r^2} + O\left(
\frac{1}{r^3}
\right).
\end{equation}
Since the Lagrangian vanishes in the limit $r\rightarrow\infty$, we see that the action integral constructed out of the renormalized action is free of IR-divergences. The corresponding non-vanishing components of the superpotential in the limit $r\rightarrow\infty$ are 
\begin{equation}
h S_t^{\ tr}=-h S_t^{\ rt}=2M\sin\theta, \quad \quad
h S_\phi^{\ tr}=-h S_\phi^{\ rt}=-3aM\sin^3\theta,
\end{equation}
which automatically lead to the physical energy-momentum four-vector
\begin{equation}
P_\mu=(M,0,0,-aM).
\end{equation}

\section{Some comments on the pure tetrad teleparallel gravity}
\label{PureTetra}
Let us mention here that there exists another approach to teleparallel gravity that can be considered to be a different theory called {\em pure tetrad teleparallel gravity}, obtained by replacing torsion with the coefficient of anholonomy \cite{Cho:1975dh,Maluf:2013gaa}. Its name stems from the fact that the only variable in this theory is the tetrad, with the spin connection set to zero in all reference frames. There is a problem with this procedure though. The point is that, as can be seen from the decomposition (\ref{gaugetet}), a spin connection is always hidden in the tetrad, which is usually overlooked in this theory. In fact, the spin connection is usually set to zero in all covariant derivatives, but not in the tetrad. Among other consequences, this procedure leads to the breaking of local Lorentz symmetry.

To understand the relation with the covariant formulation of teleparallel gravity used in this paper, we can recall that in Section~\ref{sec6}, we have used the vanishing spin connection to solve the field equations. However, in our theory this was just a necessary mid-step, and then the appropriate spin connection was calculated.  It is important to mention that for some applications, like the determination of the metric tensor, this can be sufficient, since the metric tensor is independent from the spin connection.  

As an additional comment, let us mention that in our notation the gravitational action of pure tetrad teleparallel gravity is written as $\Sw(h^a_{\ \mu},0)$. Since inertial effects were not entirely removed in the sense that the spin connection inside the tetrad has not been set to zero, such action thus represent both gravitational and inertial effects. Consequently, the energy-momentum derived from it is contaminated by inertial effects, and hence is in general divergent. A regularization process using the reference tetrads has already been developed \cite{Maluf:2005sr,Maluf:2007qq}, which yields physically sensible results. However, unlike our method, the divergences are removed only quasi-locally.

\section{Summary and Conclusions}
\label{conclu}

In general relativity, inertial and gravitational effects are both included in the Levi-Civita connection, and cannot be separated. This means that the gravitational action necessarily includes both gravitational and inertial effects. This is clear from the fact that the action $\Sbol(g)$ is non-zero in Minkowski spacetime, where gravity is absent. Of course, any complex describing the energy-momentum density in this theory will also include, in addition to the contribution coming from gravitation itself, also the contribution coming from the inertial effects. It is then necessary to use a renormalization process to remove the inertial effects from the theory. An example of such process is the {background subtraction} method that removes all inertial effects from the gravitational action. Using this action one can obtain the renormalized energy-momentum density of gravitation, to the exclusion of inertial effects. One should note that in the context of general relativity this procedure is not free of problems. For example, the divergences are not removed locally from the action, but as an integral over the whole spacetime. As a consequence, the variational principle is not in general well-defined. 

On the other hand, since in teleparallel gravity it is possible to separate inertial from gravitational effects,  it turns out possible to remove the spurious inertial contributions from the theory. From a conceptual point of view, one can understand this in the following way. The basic variables in this theory are the tetrad, in which both gravitational and inertial effects appear mixed, and the spin connection, which represents only inertial effects. The crucial point is to note that the inertial effects present in any tetrad is fully determined by the teleparallel spin connection. This means that to each tetrad there is naturally associated a specific spin connection. In this paper we have provided a method for retrieving the appropriate spin connection from any tetrad.

From the viewpoint of the action, what happens is that the teleparallel action, like in the general relativity case, represents both inertia and gravitation. However, if the spin connection is appropriately chosen according to our method, the inertial contents of the tetrad are exactly cancelled by the spin connection, giving rise to an action that represents only gravitational effects. Accordingly, the computation of the gravitational energy-momentum density will automatically yield the renormalized, physically relevant result. While conceptually our method bears some resemblance to the one by Gibbons-Hawking, or those obtained in the context of pure tetrad theory, there is an important difference: in the teleparallel method presented here the divergences coming from the inertial effects are removed locally at each point of spacetime---and not quasi-locally. As a consequence, the variational principle in the teleparallel case remains always well-defined. 

Let us conclude with the remark that our method cannot be applied straightforwardly  in the presence of a cosmological constant. This is due to the fact that the reference tetrad is defined by setting the gravitational constant to zero (see Eq.~(\ref{reftet})), but this does not switch-off the effect of the cosmological constant, which has a divergent contribution to the action on its own.  However, preliminary results show that it is possible to modify our method to address  the problem of a cosmological constant as well \cite{KAdS}. 

\section{Acknowledgements}
The authors would like to thank S. Faci for useful comments. This work was supported by FAPESP, CNPq and CAPES.



\begin{thebibliography}{10}
\addcontentsline{toc}{section}{References}


\bibitem{Landau:1982dva}
  L.~D.~Landau and E.~M.~Lifschitz,
  {\it The Classical Theory of Fields}, 4th English edition (Butterworth-Heinemann, Oxford, 1975).


\bibitem{York:1972sj}
  J.~W.~York Jr.,
  Phys.\ Rev.\ Lett.\  {\bf 28}, 1082 (1972).

\bibitem{Gibbons:1976ue}
  G.~W.~Gibbons and S.~W.~Hawking,
  Phys.\ Rev.\ D {\bf 15}, 2752 (1977).

\bibitem{Regge:1974zd}
  T.~Regge and C.~Teitelboim,
  Ann. Phys.\ (NY)  {\bf 88}, 286 (1974).

\bibitem{Hawking:1995fd}
  S.~W.~Hawking and G.~T.~Horowitz,
  Class.\ Quant.\ Grav.\  {\bf 13}, 1487 (1996)
  [arXiv:gr-qc/9501014].

\bibitem{Brown:1992br}
  J.~D.~Brown and J.~W.~York, Jr.,
  Phys.\ Rev.\ D {\bf 47}, 1407 (1993)
  [gr-qc/9209012].


\bibitem{Mann:2005yr}
  R.~B.~Mann and D.~Marolf,
  Class.\ Quant.\ Grav.\  {\bf 23}, 2927 (2006)
  [arXiv:hep-th/0511096].


\bibitem{Aldrovandi:2013wha}
  R.~Aldrovandi and J.~G.~Pereira,
  {\it Teleparallel Gravity: An Introduction},
(Springer, Dordrecht, 2012).  


\bibitem{KoNu}
S. Kobayashi and K. Nomizu, {\em Foundations of Differential Geometry}, 2nd edition (Wiley-Intersciense, New York, 1996).


\bibitem{Obukhov:2002tm}
  Y.~N.~Obukhov and J.~G.~Pereira,
  Phys.\ Rev.\ D {\bf 67} (2003) 044016
  [gr-qc/0212080].


\bibitem{HB}
M. Blagojevic and F. Hehl (eds.), {\em Gauge Theories of Gravitation: A Reader with Commentaries}, (Imperial College Press, London, 2013), see Chapter 6.


\bibitem{Weinberg}
Weinberg, S.: \textit{Gravitation and Cosmology: Principles and Applications of the General Theory of Relativity}, (Wiley, New York, 1972).


\bibitem{Obukhov:2006sk}
  Y.~N.~Obukhov and G.~F.~Rubilar,
  Phys.\ Rev.\ D {\bf 73}, 124017 (2006)
  [arXiv:gr-qc/0605045].

\bibitem{Hayashi:1967se}
  K.~Hayashi and T.~Nakano,
  Prog.\ Theor.\ Phys.\  {\bf 38}, 491 (1967).

 
\bibitem{Lucas:2009nq}
  T.~G.~Lucas, Y.~N.~Obukhov and J.~G.~Pereira,
  Phys.\ Rev.\ D {\bf 80}, 064043 (2009)
  [arXiv:0909.2418].

  
\bibitem{Zet:2003aa}
  G.~Zet,
  [arXiv:gr-qc/0308078].


\bibitem{Cho:1975dh}
  Y.~M.~Cho,
  Phys.\ Rev.\ D {\bf 14}, 2521 (1976).

\bibitem{Maluf:2013gaa}
  J.~W.~Maluf,
  Ann. Phys.\ (Berlin)  {\bf 525}, 339 (2013)
  [arXiv:1303.3897].


\bibitem{Maluf:2005sr}
  J.~W.~Maluf, M.~V.~O.~Veiga and J.~F.~da Rocha-Neto,
  Gen.\ Rel.\ Grav.\  {\bf 39}, 227 (2007)
  [arXiv:gr-qc/0507122].


\bibitem{Maluf:2007qq}
  J.~W.~Maluf, F.~F.~Faria and S.~C.~Ulhoa,
  Class.\ Quant.\ Grav.\  {\bf 24}, 2743 (2007)
  [arXiv:0704.0986].

\bibitem{KAdS}
Martin Kr\v{s}\v{s}\'ak, in preparation. 

\end{thebibliography}
\end{document}